\documentclass[review,sort&compress,2013,12pt,5p,times,twocolumn] {elsarticle} 
\usepackage[pdftex,bookmarks,colorlinks,breaklinks]{hyperref}  
\usepackage{lineno}

\journal{Nuclear Instruments and Methods in Physics Research}
\begin{document}
\begin{frontmatter}
\title{ Performance study of wavelength shifting acrylic plastic for Cherenkov light detection}
\begin{centering}
\author[focal1]{\bf B. Beckford\corref{cor1}}
\ead{beckford@aps.org}

\author[rvt2]{A. De la Puente} 
\author[focal]{  Y. Fujii}
\author[focal]{ O. Hashimoto\corref{cor2}}
\cortext[cor2]{Deceased}
\author[focal]{ M. Kaneta}
\author[focal]{ H. Kanda}
\author[focal]{ K. Maeda} 
\author[focal]{ A. Matsumura} 
\author[focal]{ S. N. Nakamura} 
\author[rvt]{N. Perez}
\author[rvt]{J. Reinhold}
\author[els]{L. Tang} 
\author[focal]{  K. Tsukada} 
\end{centering}

 \address[focal1]{American Physical Society, One Physics Ellipse, College Park, MD, 20740, USA}
 \address[focal]{Department of Physics, Tohoku University, Sendai, 980-8578, Japan}
 \address[rvt]{Department of Physics, Florida International University, Miami, Fl,33199, USA}
\address[els]{Department of Physics, Hampton University, Hampton, VA 23668, USA}
\address[rvt2]{TRIUMF Laboratory, 4004 Wesbrook Mall, Vancouver, BC V6T 2A3, Canada}

\begin{abstract}
The collection efficiency for Cherenkov light incident on a wavelength shifting plate (WLS) has been determined during a beam test at the Proton Synchrotron facility located in the  National Laboratory for High Energy Physics (KEK), Tsukuba, Japan. The experiment was conducted in order to determine the detector's response to photoelectrons converted from photons produced by a fused silica radiator; this allows for an approximation of the detector's quality.  The yield of the photoelectrons produced through internally generated Cherenkov light as well as light incident from the radiator was measured as a function of the momentum of the incident hadron beam. The yield is proportional to  sin$^2$$\theta_c$, where $\theta_{c}$ is the opening angle of the Cherenkov light created. Based on estimations and results from similarly conducted tests,  where the collection efficiency was roughly $39\%$,  the experimental result was expected to be around $40\%$ for internally produced light from the WLS. The results of the experiment determined  the photon collection response efficiency of the WLS to be roughly $62\%$ for photons created in a fused silica radiator and $41\%$ for light created in the WLS. 
\end{abstract}


\end {frontmatter}

\section{Introduction}
The use of Cherenkov detectors is now commonplace in the field of nuclear and particle physics. They have been implemented mainly for their particle identification abilities. The number of photons produced per unit energy and path length, for a particle z, is\cite{Tamm,Particle_data_book,Jackson1998}

\begin{equation}
\frac{d^{2} N} {dEdx} =  \frac {2\pi\alpha z^{2}}{h c}\sin^2\theta_c
\end{equation}

in terms of wavelength it may be written as:

\begin{equation}
\frac{d^{2} N} {d\lambda dx} =  \frac {2\pi\alpha^{2} z^{2}}{\lambda^2}\sin^2\theta_c
\end{equation}

\begin{figure}[t]
\begin{center}
\includegraphics*[width=1.0\columnwidth]{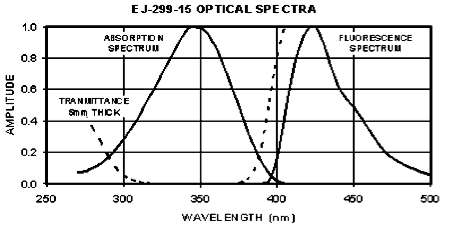}
\end{center}
\vspace{-0.5cm}
 \caption{Wavelength shifting plastic's absorption and flouresence spectrum. \cite{elgen}}
\label{wls_spectrum} 
\end{figure}

Cherenkov detectors are known for requiring a high efficiency in detection and acquisition of photons in the ultra-violet range of the electromagnetic spectrum. The distribution of the Cherenkov radiation follows $dN/d\lambda \propto 1/\lambda^2$ as implied in equation 2 \cite{Jackson1998}.The number of photoelectrons, produced in a Cherenkov radiator per unit of length \emph{L} , that will be detected by a PMT is given by: 

\begin{equation} 
{dN_{p.e.}}=370eV^{-1}cm^{-1}L\int{\epsilon_{coll}(E)sin^2\theta_c(E)\mathrm{d}E}
\end{equation}

where,  both $\epsilon(E)$, the efficiency for collecting Cherenkov light and converting it into photoelectrons, and sin$^{2}$ $\theta_c(E)$ are functions of energy, and \emph{L} is the thickness of the radiator. \cite{Particle_data_book}.  If the previous equation is intergrated over all photon energy ranges such that:

\begin{equation}
{N_o=370eV^{-1}cm^{-1}\int{\epsilon_{coll}(E)}\mathrm{d}E}
\end{equation}

an approximation of the total number of photoelectrons detected can be written as: 

\begin{equation}
N_{p.e.} \approxeq LN_0\langle \sin^2\theta_c\rangle
\end{equation}

where, $\langle$sin$^{2}$$\theta_c$$\rangle$ is the average squared opening Cherenkov angle, for the photon energy range, and $N_{o}$ is the Cherenkov detector's quality \emph{$N_o$}, also referred  to as  it's \emph{figure of merit}. Therefore, the collection efficiency of the produced Cherenkov radiation is dependent on the sensitivity of the detector to light in the UV range. Most commonly produced detectors using photomultipliers are known to have collection efficiencies of in the range of  $20-25\%$ at 400$-$420 nm wavelengths. A stratagem that has been adopted is the use of wavelength shifter to bolster the light yield. The principle lies in using the wavelength shifter, an organic chemical additives that converts the wavelength of  absorbed photons from a shorter to a longer wavelength. The additive has a high absorption ability around 350 nm, as shown in Figure~\ref{wls_spectrum}, where the amplitude of absorption and fluorescence as a functions of wavelengths are shown.  Evidently, the absorbed light will be reemitted at a longer wavelength, hence achieving a match between the wavelength shifter peak emission with that of the photomultiplier (PMT). However, an over abundance in the addition of the wavelength shifter can result in a loss of the signal light as a result of attenuation.

\subsection{Motivation}
In this paper, we studied the photon detection capability (\emph{N$_{o}$}) of an acrylic plastic detector that used a WLS additive. It is widely known that the yield of Cherenkov detectors are readily improved with the addition of a WLS. This work is a continuation of a previous study conducted using aerogel as a radiator and WLS as a collector. The collection efficiency of Cherenkov light created within the aerogel furnished poor results due to scattering within the aerogel, yet still be indicated that the collection efficiency was for  UV photons to approximately $41\%$~\cite{Carl2004}.  There have been advances in other detectors that are composed of aerogel, and wavelength shifter such as the one developed by Novosibirsk~\cite{Novo}. Other efforts have shown an increase in light yield of a factor of roughly 4 for water Cherenkov detectors~\cite{SNO}. Extensive investigations have been performed on different variations of fused silica as Cherenkov radiators in order to determine their optical properties and radiation hardness~\cite{Dirc}. As such, an evaluation of WLS collection capability was performed with a synthetic fused silica as a more defined radiator~\cite{edmund}. The results of this experiment will clarify some unanswered questions and will be used to improve upon or decide whether a proposed one-dimensional RICH detector, that uses wave-length shifter bars, would be more efficient and can be used to replace a conventional RICH detector.

\section{Experimental Setup}
\subsection{Detector Design}
A test experiment was designed to ascertain the usefulness of  a detector that incorporates a WLS plastic. The detector consisted of a synthetically made fused silica (FS) radiator that was placed 15.0 cm from the wavelength shifter.

\begin{figure}[t]
\begin{center}\includegraphics*[width=1.0\columnwidth]{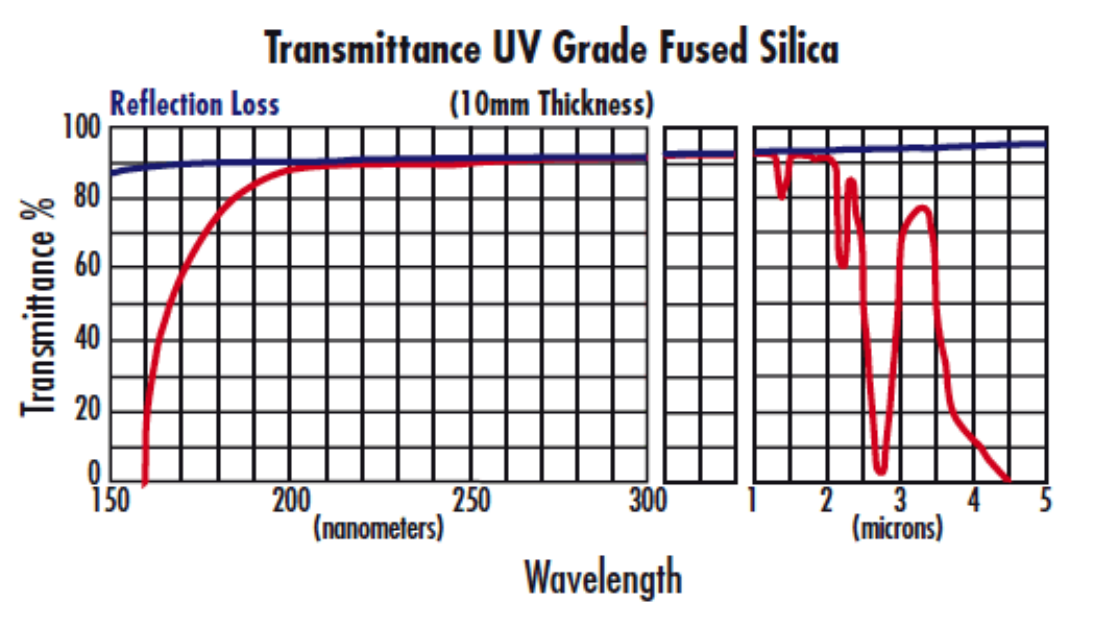}\end{center}
\vspace{-0.7cm}
 \caption{Transmittance spectrum of UV grade fused silica.  \cite{edmund} }
\label{transmittance}
\end{figure}

The radiator had an index of refraction ($\emph{n}_{r}$) $=$ 1.476 correspondening to a wavelength of  $\lambda$ = 351 nm, 
and the physical dimensions of 50 x 50 x 8 mm$^3$. A complete listing of the fused siclia's index of refraction as a function of wavelength is presented in Table~\ref{index_of_refraction_table}.

\begin{table}[ht]
\caption{ Fused silica refractive index.~\cite{Optics} }
\begin{center}
\begin{tabular}{||c|c||}
  \hline
  \hline
  Wavelength($nm$)&Index of refraction(\emph{n})\\

  \hline

  296.7 	& 	1.48873 	\\
  
  302.2 	& 	1.48719 	\\
  
  330.3 	& 	1.48054 	\\
  
  340.4 	& 	1.47858	\\
  
  351.1 	& 	1.47671 	\\
  
  361.1 	& 	1.47513 	\\
  
  365.0 	& 	1.47454 	\\
  
  404.7 	& 	1.46962	\\

  435.8 	& 	1.46669	\\ 
  
  441.6 	& 	1.46622  \\
  
   457.9	& 	1.46498  \\ 
  
   476.5 	& 	1.46372\\
  
   486.1	& 	1.46313  \\
   
   496.5	& 	1.46252\\
   
   514.5 	& 	1.46156\\
   
   532.0	& 	1.46071\\
   
   546.1	& 	1.46008\\  
  
  \hline
  \hline
\end{tabular}
\end{center}
\label{index_of_refraction_table}
\end{table}

The acrylic plastic bar was 1.27 x 10 x 30.5 cm$^3$ in size and contained a wavelength shifting flourescent additive~\cite{elgen}. The transmittance of the radiator is shown in Figure~\ref{transmittance}.

\begin{figure}[ht]
\begin{center}
\includegraphics*[width=1.0\columnwidth]{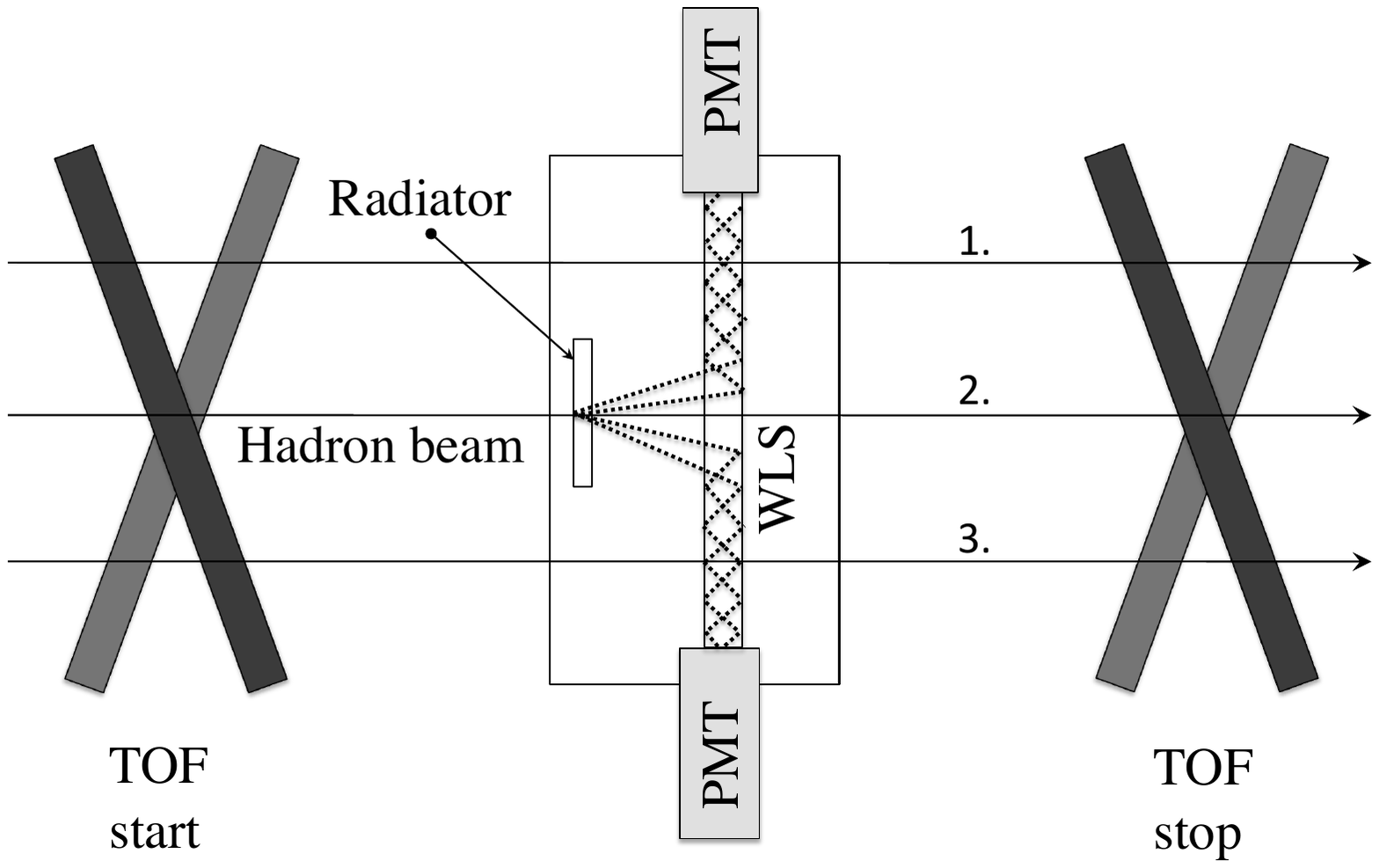}
\end{center}
\vspace{-0.7cm} 
\caption{Schematic view of the Cherenkov detector setup along hadron beam line. The three orientations in which the beam transversed the detector are shown by the solid black lines. }
\label{experimental_setup} 
\end{figure}

The light that falls upon the wavelength shifter in addition to the Cherenkov light created in the bar was converted to a longer wavelength and was isotropically re-emitted~\cite{Carl2004}. The peak of the flourescence spectrum, 400 - 450 nm, lies in the peak effective range of the photomultiplier. Having and index of refraction of $\emph{n}_{wls}$ =1.49, the critical angle for total internal reflection is $42^0$ for the WLS. The estimated loss of light that is emitted below the critical angle is roughly 51$\%$. The quantum efficiency of the WLS is in the range of $84\%$, therefore, 41$\%$ of the light isotropically emitted should reflect along the length of the wavelength shifting bar towards the attached photomultipliers.  

A pair of 12.5 cm diameter photomultipliers (PMTS) (XP457B/D1)~\cite{photonis} were glued to both ends of the wavelength shifter with optical cement (EJ-500)~\cite{elgen_cement}. The radiator was held in position using black foam core board and the entire setup was contained within a light tight box. The implementation of the black foam core along with covering the interiors surfaces of the detector was undertaken to prevent any re-scattering of photons within the detector housing. 

\subsection{Method}

\begin{figure}[h!]
\begin{center}\includegraphics*[width=1.0\columnwidth]{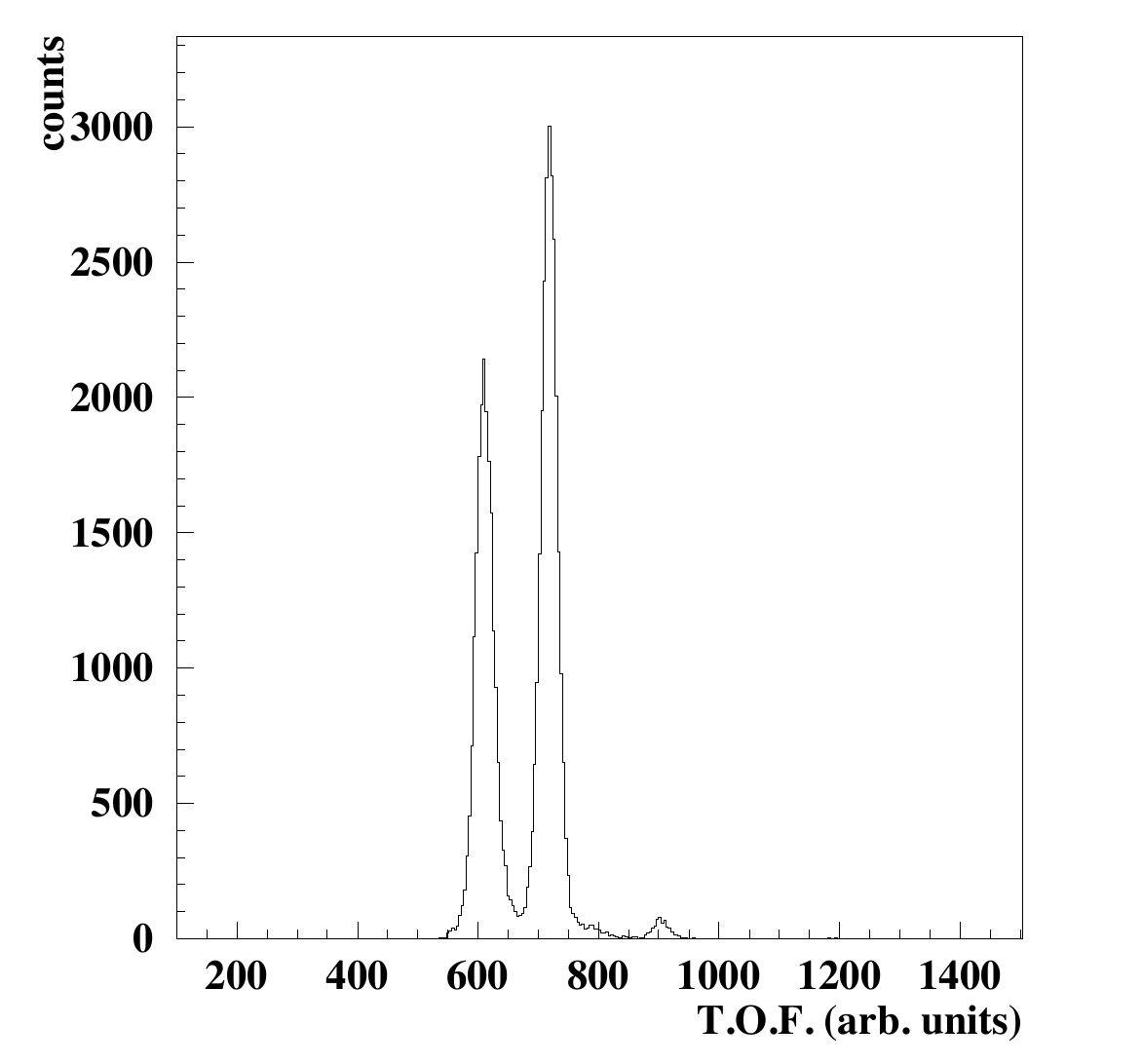}
\end{center}
\vspace{-0.5cm} 
\caption{ Time-of-Flight (TOF) distribution for pions, protons, and deuterons, left to right respectively, at 1.5 GeV/c momentum.}
\label{TOF}
\end{figure}

The detector was placed along one of the secondary hadron beam line of the KEK facility. The secondary beam provided a mixed hadron beam consisting of protons, kaons, pions, and deuterons with momentum up to 2 GeV/c. Figure~\ref{experimental_setup} shows the experimental detector arrangement along the beam line. 
The beam was extracted at several momenta; the momenta selected for the detector test were 1.05, 1.2, 1.35, 1.5 and 1.7 GeV/c.

 The Cherenkov opening angle can be determined from the relation:
\begin{equation}
\cos\theta_c = \frac{1}{\beta n}
\end{equation}

and, thus, the opening angle for a particle whose velocity $\beta$ = \emph{v/c}, where $\beta$ is the ratio of a particles velocity in the medium to that of light in a vacuum, equal to 1 is $52.6^0$. There were two small crossed TOF scintillators placed at the beginning and end of the beam line. The initial orientation was such that the beam was incident on the radiator as well as the WLS. This is illustrated by position 2 in Figure~\ref{experimental_setup}.
Cherenkov light can also be produced as the hadron beam transversed the WLS. Therefore, the detector was  moved to an orientation in which the beam would be directed only through the WLS and not the radiator. This was performed for both positions left and right of the radiator. Movement of the detector position allowed the calculation of the light output of the radiator from the difference between the light output gathered with beam center and that of the average of the left and right positions, where the Cherenkov light yielded was purely from the WLS.  Positions 1 and 3 in Figure~\ref{experimental_setup} displays these orientations.

\section{Data Analysis}

\begin{figure}[ht]
\begin{center}
\includegraphics*[width=1.07\columnwidth]{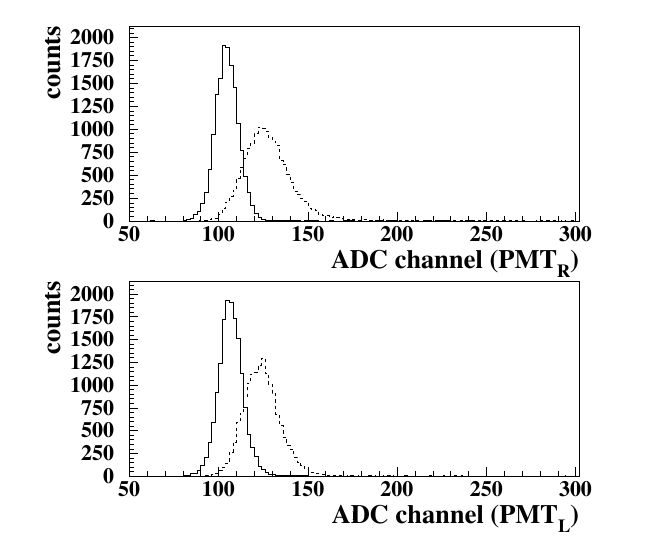}
\end{center}
\vspace{-0.5cm} 
\caption{Uncalibrated superposition of the pedestal and first photoelectron peak. The upper figure is the response of the right PMT,  the lower is the response of the left PMT, with respect to the beam line.}
\label{ped_peak} 
\end{figure}

\begin{table}
 \caption{ Photoelectron yield for protons at various beam momenta} 
\scriptsize
\begin{center}
\begin{tabular}{||c|c|c|c|c||}
  \hline
  \hline
  \emph{P}($GeV/c$) 	& N.p.e(Cntr.) 		& N.p.e(L) 			& N.p.e(R) 			& N.p.e(Rad.) \\
  \hline
  		1.05 		& 23.9$\pm$1.8 	& 15.1$\pm$2.2 	& 14.1$\pm$1.8 	& 9.3 $\pm$2.1 \\
  
  		1.20 		& 32.5$\pm$1.2 	& 21.2$\pm$1.1 	& 19.4$\pm$1.1 	& 13.1$\pm$1.4 \\
  
  		1.35 		& 38.9$\pm$0.8 	& 24.4$\pm$1.0 	& 22.8$\pm$1.3 	& 17.3$\pm$1.4 \\
  
  		1.50 		& 44.8$\pm$1.1 	& 25.6$\pm$1.8 	& 26.7$\pm$1.2 	& 18.1$\pm$1.6 \\
  
  		1.70 		& 46.4$\pm$1.6 	& 31.2$\pm$1.8 	& 29.2$\pm$2.4 	& 17.5$\pm$1.7 \\
  \hline
  \hline
\end{tabular}
\end{center}
\label{photo_yield_table}
\end{table}

The time of flight distribution was a necessary part of the data analysis technique. It allowed for the proper timing cuts to be made in order to separate the light generated by each type of particle. Figure~\ref{TOF} illustrates a typical TOF distribution for pions, protons, and deuterons, left to right respectively, at 1.5 GeV/c momentum. Using the determined timing conditions the appropriate cuts on the raw ADC spectra were performed. The hadron beam consisted of kaons protons,  deuterons and pions , however, the Cherenkov light that would be created from pions and kaons as they transversed the fused silica would be be emitted above the the critical angle for internal reflection, and would subsequently be confined within the radiator. The light collected was from chiefly pions and protons,  but the analysis focused on the light produced from protons. The raw ADC spectra were calibrated to show the number of photoelectrons by determining the ADC pedestal channel position and ADC channel position of the one photoelectron  distribution. The difference between the pedestal and the position of the one photoelectron distribution translates into a proportionality between the collected charge at the anode and the number of collected photoelectrons.

\begin{figure}[ht]
\begin{center}\includegraphics*[width=1.0\columnwidth]{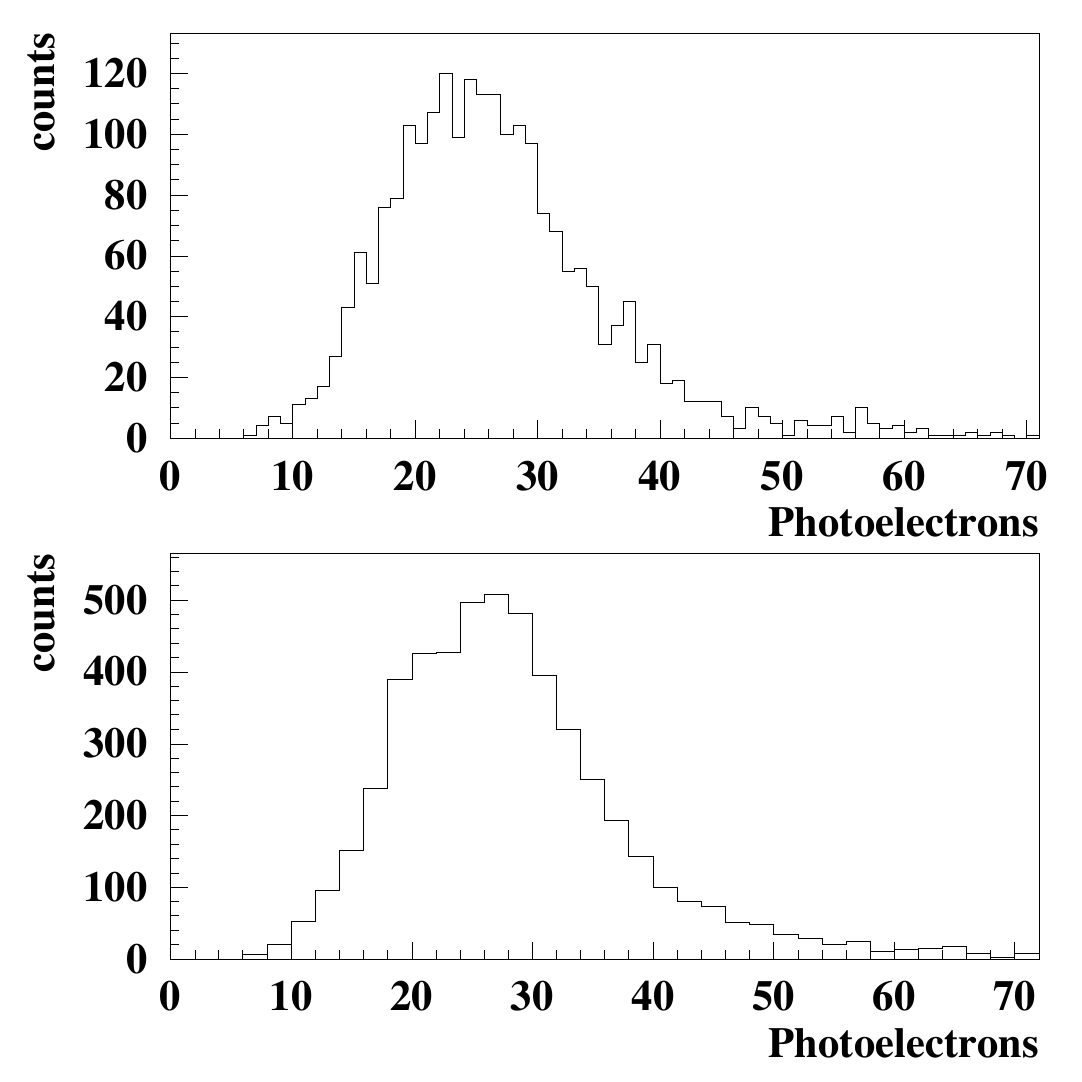}\end{center}
\vspace{-.5cm} 
\caption{Calibrated photoelectron yield for protons at 1.5 GeV/c for detector orientations. The upper figure is for the beam right position,  the lower is for the beam left position. The number of photoelectrons collected was 25.6$\pm{ 1.8}$ and 26.7$\pm {1.2}$ for the left and right orientations respectively.}
\label{npe_histograms} 
\end{figure}

An uncalibrated overlay of the pedestal and first photoelectron peak is illustrated in Figure~\ref{ped_peak}. Data were obtained by triggering the data acquisition randomly and with low thresholds on the individual PMTs. The photoelectron production of the radiator at each momentum was determined by subtracting the average photoelectron yield of beam left and right, from the photoelectron yield at the beam center position. Where, at the  beam center position the number of photoelectrons detected is a combination of the amount generated by the FS and the WLS. The calibrated photoelectron yield for protons at 1.5 GeV/c beam left and right orientations are given in Figure~\ref{npe_histograms} In the figure the upper panel is the beam right position,  the lower is the beam left position. The number of photoelectrons collected was 25.6$\pm{ 1.8}$ and 26.7$\pm {1.2}$ for the left and right orientations respectively. 

\section{Results}
\begin{figure} [h!] 
\begin{center}
\includegraphics*[width=1.01\columnwidth]{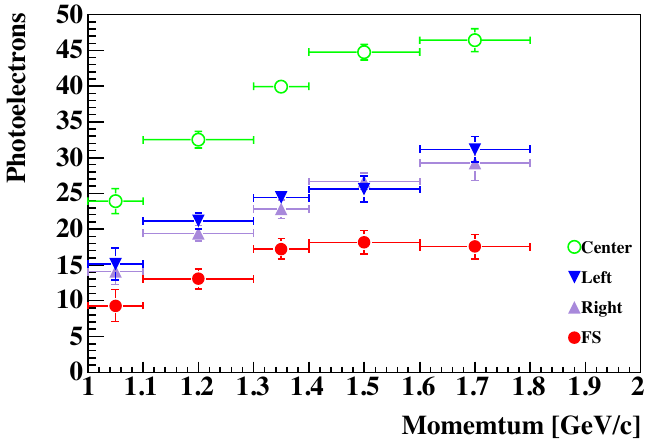}
\end{center}
\vspace{-0.5cm} 
\caption{Number of photoelectrons as a function of momentum [GeV/c] created in the WLS and fused silica. Data for WLS and fused silica (open circle), WLS (triangles), and fused silica (FS) (solid circle) are plotted. }
\label{npe_vs_momentum} 
\end{figure}

The number of Cherenkov photons detected for each momenta and detector orientation are listed in Table~\ref{photo_yield_table}. In these results presented in this study only the statistical error are reported. In Figure~\ref{npe_vs_momentum} the internally created light by the WLS and radiator, the WLS for the hadron beam transiting left and right, and externally produced light of the fused silica radiator as a function of momentum are shown as open circles, solid triangles,  and solid circles respectively. 

\begin{figure}[h!]
\begin{center}
\includegraphics*[width=1.01\columnwidth]{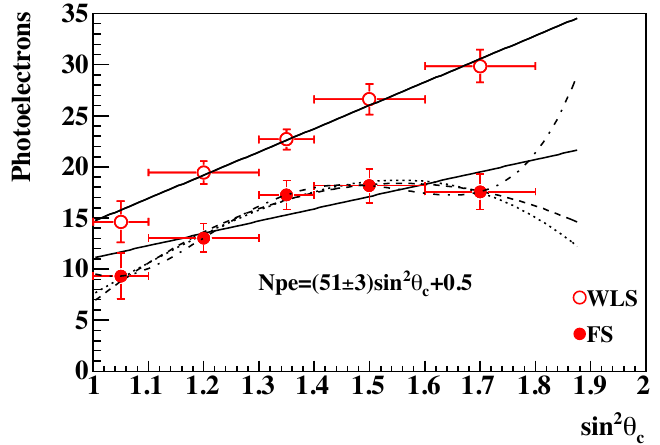}
\end{center}
\vspace{-0.5cm} 
\caption{ Number of  photoelectrons as a function of $\sin^2\theta_c$ for the WLS (open cirlces) and fused silica (FS) (solid circle) responses are plotted. Linear, 2$^{nd}, $3$^{rd}$, and 4$^{th}$ degree polynomial fits are shown, where $\epsilon_{coll}$ and the index of refraction for the radiator $\emph {n}_{r}$ are both free parameters}
\label{npe_vs_sin}
\end{figure}

The data were fitted to a linear dependence of $\emph{N$_{p.e}$}\approxeq LN_0  $sin $^2\theta_{c}$, where the radiator index of refraction $\emph {n}_{r}$ is 1.476, for corresponding  wavelength of  $\lambda$ = 351 nm and $\emph{${N_0}$}$ is a free parameter, a determination of the quality of the detector, and \emph{L }is the thickness of the radiator.  As shown, the last data point  does not follow the increasing trend of the other data points, indicating that the number of photons collected was reduced due to partial internal reflection of photons within the radiator above 1.5 GeV/c  or as a consequence of photon loss that may have occurred between the air and WLS boundaries. The normalization of the slope to the radiator thickness of 0.8 cm, yielded a value of \emph{${N_o}$} = 63.86, this would in turn, correspond to a collection efficiency of $64\%$ for external light produced by the radiator. The number of photoelectron detected that were produced internally by the WLS itself, was calculated to be $41\%$.

In Figure~\ref{npe_vs_sin} the number of photoelectrons detected as a function of the Cherenkov angles are given. If both  $\epsilon_{coll}$ and the index of refraction for the radiator $\emph {n}_{r}$ are left as free parameters for fitting the experimental data, as shown in Figure~\ref{npe_vs_sin} by the dotted curve, the calculated collection efficiency is approximately $61.5\%$.

\subsection{Discussion}
A wavelength-shifting Cherenkov detector was developed and a test experiment was performed at the Proton Synchrotron facility located in the  National Laboratory for High Energy Physics (KEK), Tsukuba, Japan. For Cherenkov photons incident on the WLS, a collection efficiency of $64\%$ was determined.  
Considering the results of both fitting procedures, the collection efficiency can be concluded to be between $61-64\%$ for Cherenkov photons created in the fused silica. These values are comparable to more traditional means of detecting photons in RICH detectors, like CsI photocathodes or large PMT arrays. Therefore, we are currently investigating the feasibility of using arrays of WLS bars as a means of constructing cost efficient one-dimensional RICH detectors. The design considers the use of an array of plastic bars with a wavelength shifter substance. These bars are read out by photomultipliers to map the Cherenkov cones in one dimension instead of the two-dimensional mapping of conventional RICH. The main goal is to reduce the number of PMT and the associated electronics, which in turn, reduces construction and operational costs.

\subsection{Acknowledgements}
The authors would like to thank the staff of KEK accelerator facility. And we would also like to thank the faculty and students of Florida International University and Tohoku University for their collaboration on this project.

\section{References}
 \bibliographystyle{model1b-num-names}

\begin{thebibliography}{1}

\bibitem{Tamm}I.E. Tamm ,et al., Coherent Radiation of Fast Electrons in a Medium SSSR, 14 107 (1937)
\bibitem{Particle_data_book}B.N. Ratcliff, Particle Physics Booklet. (2004)
\bibitem{Jackson1998}J.D. Jackson,Classical Electrodynamics, 3rd Edition. Wiley, New York, 1998 
\bibitem{Carl2004}M. Carl, et al., Nuclear Instr. And Methods. A 527 (2004)
\bibitem{Novo}A.Yu. Barnyakov, et al., Nuclear Instr. And Methods. A 553 (2005)
\bibitem{SNO}X. Dai, et al., Nuclear Instr. And Methods. A 589 (2008) 
\bibitem{Dirc}J. Cohen-Tanugi, et al., Nuclear Instr. And Methods. A 515 (2003)
\bibitem{edmund}Edmund Optics Inc.101 East Gloucester Pike, Barrington, NJ 08007-1380 USA
\bibitem{Optics}Melles Griot Optics Guide http://www.mellesgriot.com/pdf/CatalogX/X-04-11- 13.pdf
\bibitem{elgen}Elgen Technology,PO BOx 870, 300 Crane Street, Sweetwater,TX 79550, USA. 
\bibitem{photonis}Photonis Imaging Sensors, Avenue Roger Roncier. Z.I. Beauregard, B.P. 520 19106 Brive Cedex, France.
\bibitem{elgen_cement}Eljen Technology, Apace Science Inc, Mitomo-Bldg. 2-6F, 2-8, Ebisu-Nishi 2-Chome, Shibuya-Ku, Tokyo, Japan 150-0021

\end{thebibliography}

\end{document}